\documentclass[12pt]{iopart}

\usepackage[utf8]{inputenc}
\usepackage{comment}
\usepackage{multirow}
\usepackage[english]{babel}
\usepackage{iopams}
\usepackage{graphicx}
\expandafter\let\csname equation*\endcsname\relax
\expandafter\let\csname endequation*\endcsname\relax
\usepackage{amsmath}

\usepackage{color}
\usepackage[colorinlistoftodos]{todonotes}
\usepackage{float}
\usepackage{textcomp}
\usepackage{subfig}

\begin{document}

\title[Diagram Reduction in Critical Dynamics]{Diagram Reduction in Problem of Critical Dynamics of Ferromagnets: 4-Loop Approximation}

\date{}

\author{L.\,Ts. Adzhemyan$^1$, E.\,V. Ivanova$^1$, M.\,V. Kompaniets$^1$ and  S.\,Ye. Vorobyeva$^1$}
\address{$^1$  Saint-Petersburg State University,  7-9 Universitetskaya nab. Saint-Petersburg, Russian Federation}
\eads{\mailto{l.adzhemyan@spbu.ru},  \mailto{ellekspb@gmail.com}, \mailto{m.kompaniets@spbu.ru}
\mailto{svetlana.e.vorobyeva@gmail.com},}
\begin{abstract}
Within the framework of the renormalization group approach to the models of critical dynamics, we propose a method for a considerable reduction of the number of integrals needed to calculate the critical exponents. With this method we perform a calculation of the critical exponent $z$ of model A at 4-loop level, where our method allows to reduce number of integrals from 66 to 17. The way of constructing the integrand in Feynman representation of such diagrams is discussed.  Integrals were estimated numerically with Sector Decomposition technique.
\\

\end{abstract}
\vspace{2pc}
\noindent{\it Keywords}: renormalization group, $\varepsilon$-expansion, multi-loop diagrams, critical exponents

\maketitle 

\section{Introduction}
Model A of critical dynamics describes critical slowing down effect for systems with non-conserved order parameter~\cite{obzor1,obzor2,Vasiliev}. Usually this model is used as a theoretical model for critical behavior of ferromagnets~\cite{amodelexp1}.
Recently the new classes of materials were investigated and it was found that this model describes phase transitions in multiferroics~\cite{amodelexp3} and in the systems with ordering phase transitions~\cite{amodelexp2} as well. 
An additional motivation to study model A is also the fact that it is the simplest model of critical dynamics and new technical methods can be tested on it.



Despite the fact that renormalization group is one of the well acknowledged theoretical methods for investigation of continuous phase transitions, 
the application of this method to the problems of critical dynamics faces much greater difficulties in comparison with the problems of critical statics. The analytic results obtained here are limited in the best case to the third order of perturbation theory~\cite{2}, whereas in the static  $\varphi^4$ theory the six-loop result \cite{chet, komp6, mkomp6} is currently reached, and for the anomalous dimension of the field -- seven-loop  \cite{schnetz7}. In problems of critical dynamics a noticeable lag occurs also in numerical calculations, in which the Sector Decomposition technique of the calculation of Feynman diagrams \cite{SD} proved to be very effective in critical statics problems (5 loops  and partially 6 loops in the theory $\varphi^4$ \cite{mkomp6,5loop}), while  in the dynamic problems this method has so far been used only in the two-loop approximation \cite{Emodel}.

The calculation of multi-loop diagrams in critical dynamics, taking into account their complexity, requires considerable time, thus the problem of reducing the number of calculated diagrams arises.
In this paper, a numerical calculation of the renormalization group functions of the model A \cite{obzor1,obzor2} is performed in the fourth order of perturbation theory. 
We present a method that allows to reduce significantly the number of Feynman diagrams to be calculated  (``reduction'' of diagrams) by appropriate grouping of the original diagrams of the theory. The rules  for constructing these diagrams and their integrand in the Feynman representation directly from the graph are formulated.
The subsequent numerical four-loop calculation was carried out using the Sector Decomposition method.

The paper is organized as follows:
in section \ref{sec:renorm} we recall renormalization procedure for the model A with the use of dimensional regularization ($d=4-\varepsilon$) and minimal subtraction scheme (MS). In the next section we present the diagrammatic representation for model A. In section \ref{sec:reduction} we describe the method of diagram reduction. In subsequent section we present the four loop results for the dynamical critical exponent $z$, which are followed by conclusion.
In  \ref{appendixa} we present one more example of the diagram reduction and in \ref{appendixb} we discuss the Feynman representation for diagrams in models of critical dynamics.

\section{Renormalization of the model}
\label{sec:renorm}

Non-renormalized action of model $A$ of critical dynamics in the space with the dimension $d=4-\varepsilon$  is determined by the set $\phi_0$ of two non-renormalized fields $\phi_0\equiv \{\psi_0,\,\psi'_0\}$  and has the form \cite{Vasiliev}:
\begin{eqnarray}\label{S0}
S_0(\phi_0)= \lambda_0 \psi_0' \psi_0' + \psi_0'[- \partial_t \psi_0 + \lambda_0 \delta S_0^{st}/\delta\psi_0 ]= \nonumber\\=
 \lambda_0 \psi_0' \psi_0' + \psi_0'[ -\partial_t \psi_0 + \lambda_0 (\partial^2 \psi_0 -  \tau_0 \psi_0 - \frac{1}{3!}g_0\psi_0^3)]
\end{eqnarray}
with non-renormalized static action
\begin{equation}\label{Sst0}
S_0^{st}(\phi_0)= - (\partial \psi_0)^2/2 -  \tau_0 \psi_0^2/2 - \frac{1}{4!}g_0\psi_0^4\,.
\end{equation}
Renormalized action $S_R$ obtained by multiplicative renormalization of parameters and fields can be represented as the sum $S_R=S_B+\Delta S$ of basic action $S_B$ and counterterms $\Delta S$~\cite{Vasiliev}:
 \begin{equation}\label{SB}
S_B= \lambda \psi' \psi' + \psi'[- \partial_t \psi + \lambda ( \partial^2 \psi -  \tau \psi - \frac{1}{3!}\mu^{\varepsilon}g \psi^3)]\,,
\end{equation}
\begin{equation}\label{SR}
S_R=Z_1 \lambda \psi' \psi' + \psi'[-Z_2 \partial_t \psi + \lambda (Z_3 \partial^2 \psi - Z_4 \tau \psi - \frac{1}{3!}Z_5 \mu^{\varepsilon}g \psi^3)]\,,
\end{equation}
where
\begin{equation}\label{ren}
\lambda_0=\lambda Z_{\lambda},\quad \tau_0=\tau Z_{\tau},\quad g_0=g \mu^{\varepsilon} Z_g,\quad \psi_0=\psi Z_{\psi}, \quad \psi_0'=\psi'Z_{\psi'}\,,
\end{equation}
\begin{eqnarray}\label{ren1}
Z_1=Z_{\lambda}Z_{\psi'}^2, \quad Z_2=Z_{\psi'}Z_{\psi}, \quad Z_3=Z_{\psi'}Z_{\lambda}Z_{\psi}, \\ \nonumber
Z_4=Z_{\psi'}Z_{\lambda}Z_{\tau}Z_{\psi}, \quad Z_5=Z_{\psi'}Z_{\lambda}Z_g Z_{\psi}^3\,.
\end{eqnarray}

It follows from the multiplicative renormalizability of the models (\ref{S0}), (\ref{Sst0}), that the renormalization constants $Z_\psi, Z_\tau, Z_g$ in this model coincide with the static ones (i.e. of the model (\ref{Sst0}))
\begin{equation}\label{ZZ}
Z_{\psi}=(Z_{\psi})_{st}\,,\quad Z_\tau=(Z_\tau)_{st},\,,\quad  Z_g=(Z_g)_{st}
\end{equation}
and the relation  $Z_{\psi'}Z_\lambda=Z_\psi$ fulfilled~\cite{Vasiliev}. This means that the renormalization constants $Z_3, Z_4, Z_5$ are purely static and  
\begin{equation}\label{Z12}
Z_1=Z_2.
\end{equation}\label{ren}
The only new renormalization constant is  
\begin{equation}\label{Zlambda}
Z_\lambda=Z_1^{-1}Z_\psi^2=Z_2^{-1}Z_\psi^2\,.
\end{equation}
For our purposes, it is convenient to calculate it through the renormalization constant $Z_1$, which is determined from the diagrams of the 1-irreducible function $\Gamma_{\psi'\psi'}=\langle\psi'\psi'\rangle_{1-irr}/(2\lambda)$ on the zero external frequency $\omega$ and the momentum $p$. For the expansion of this function into a perturbation theory series, we will use the coupling constant $u=\frac{S_d}{(2\pi)^d}g$, where $S_d=\frac{2\pi^{d/2}}{\Gamma(d/2)}$ is the area of the $d$-dimensional unit sphere. This expansion has the form
\begin{equation}\label{gamma}
\Gamma^R_{\psi'\psi'}|_{\omega=0, p=0}=Z_1(1+u^2 Z_g^2(\mu^2/\tau)^{\varepsilon} Z_{\tau}^{-\varepsilon }A^{(2)}+u^3 Z_g^3(\mu^2/\tau)^{3\varepsilon/2} Z_{\tau}^{-3\varepsilon/2 }A^{(3)}+...)\,.
\end{equation}

For renormalization of the model (\ref{SR}) we will use the  minimal subtraction (MS) scheme where counterterms contain only poles in $\varepsilon$.
The renormalization constants $Z_g$ and $Z_\tau$ in (\ref{gamma}) are known from the statics, while $Z_1$ at the 4-loop approximation in the MS scheme  has the following form
\begin{equation}\label{Z1}
Z_1=1+\frac{z_{21}}{\varepsilon}u^2+\left(\frac{z_{32}}{\varepsilon^2}+\frac{z_{31}}{\varepsilon}\right)u^3+\left(\frac{z_{43}}{\varepsilon^3}+\frac{z_{42}}{\varepsilon^2}+\frac{z_{41}}{\varepsilon}\right)u^4+{\cal O}(u^5)\,,
\end{equation}
coefficients $z_{nk}$ 
can be found from the condition of the absence of poles in $\varepsilon$ in the function $\Gamma^R_{\psi'\psi'}|_{\omega=0,p=0}$, thus the main technical problem is to calculate the coefficients
 $A^{(2)}, A^{(3)}, A^{(4)}$ in (\ref{gamma}).

\section{Diagrammatic representation after integration  over internal time variables }
 Propagators of the model (\ref{SB}) in the time-momentum $(t,k)$ representation have the form:  
\begin{equation}
\begin{split}
&\includegraphics[width=1.4cm]{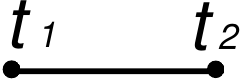} = \left\langle \psi(t_1) \psi(t_2) \right\rangle = \frac{1}{E_k} \exp^{-\lambda E_k |t_1-t_2|}\,,\\ 
&\includegraphics[width=1.4cm]{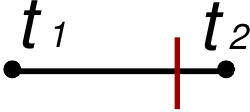} =\left\langle \psi(t_1) \psi'(t_2) \right\rangle = \theta (t_1-t_2) \exp^{-\lambda E_k(t_1-t_2)} \,,\\
&\includegraphics[width=1.4cm]{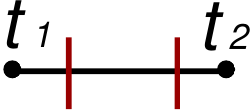} = \left\langle \psi'(t_1) \psi'(t_2) \right\rangle = 0\,, \\
&\mbox{where} \;\; E_k \equiv k^2+\tau \,.
\end{split}
\label{prop}
\end{equation}
 The simple exponential dependence of these propagators on time  makes it easy to integrate diagrams in $(t,k)$ representation over internal time variables and reduce the problem to integration over momenta (momentum representation). The result of integration over time can be expressed in a diagram language using the technique of ``time versions'' (see, for example, \cite{Vasiliev}). Let us remind this technique with the following diagram, considered at zero external frequency $\omega$:
\begin{equation} \label{example}
\begin{matrix}
\includegraphics[angle=0, width=0.2\textwidth]{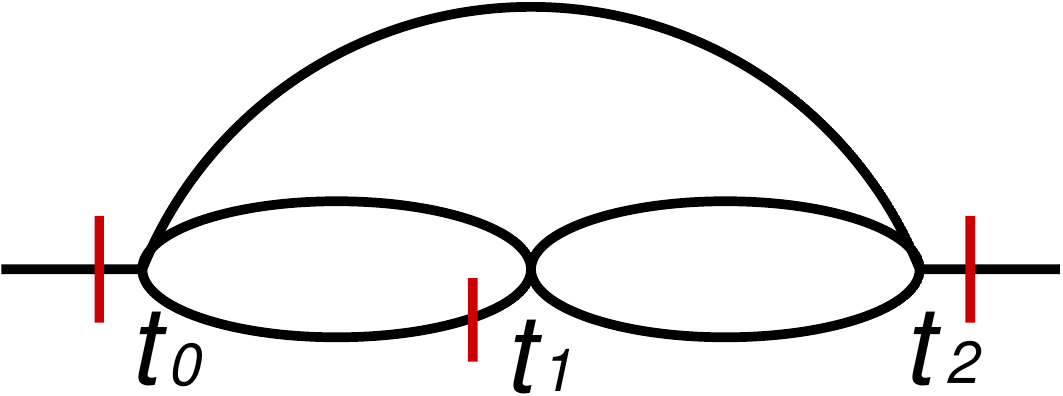}
\end{matrix}\Big|_{\omega=0}\;.
\end{equation}
 Taking into account the $\theta$--function in the propagator $\langle\psi\psi'\rangle$~(\ref{prop}), the domain of integration in (\ref{example}) can be represented as 3 contributions (time versions)
\begin{eqnarray}\label{t}
  (t_0>t_1>t_2) \quad + \quad (t_0>t_2>t_1) \quad + \quad (t_2>t_0>t_1) \,. 
\end{eqnarray}
Explicit integration over internal times for each time version (\ref{t}) can be represented with new diagrammatic technique:
\begin{equation} 
\begin{matrix}
\includegraphics[angle=0, width=0.21\textwidth]{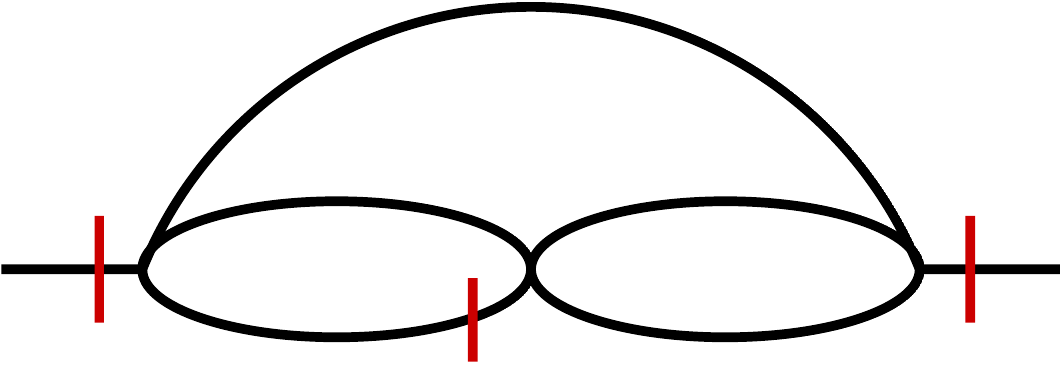}
\end{matrix} \Big|_{\omega =0}
=
\begin{matrix}
\includegraphics[angle=0, width=0.21\textwidth]{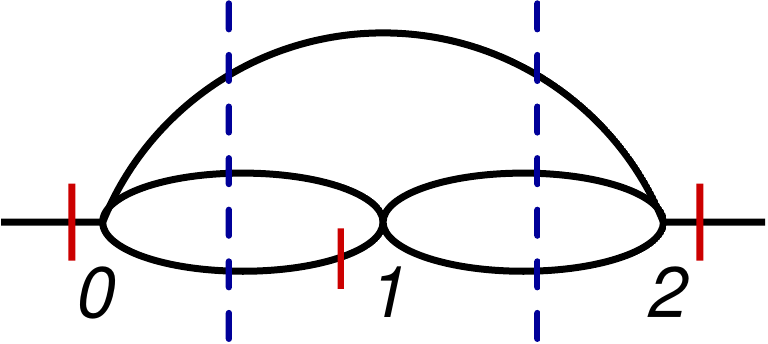}
\end{matrix}
+\begin{matrix}
\includegraphics[angle=0, width=0.16\textwidth]{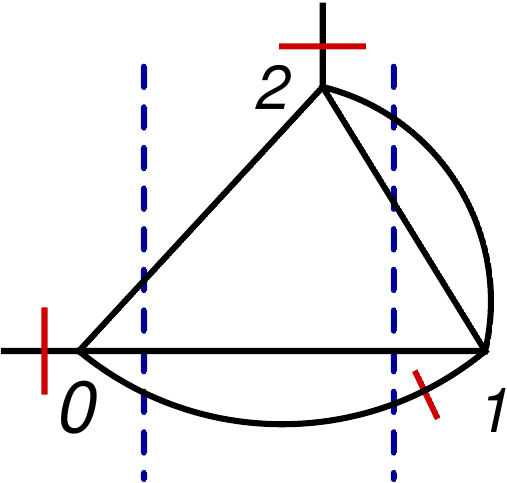}
\end{matrix}
+
\begin{matrix}
\includegraphics[angle=0, width=0.175\textwidth]{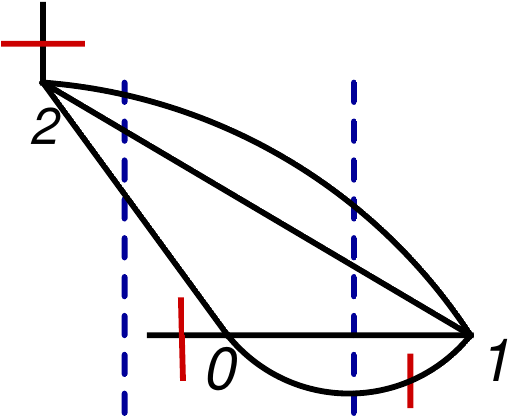}
\end{matrix} \,.
\label{tv}
\end{equation}
In this diagrammatic technique we associate factor $1/E_{k_i}$ with each solid line ($\langle\psi\psi\rangle$), factor $1$ with each solid line with dash ($\langle\psi\psi'\rangle$) and  factor $1/\sum_i E_{k_i}$ with dotted line, the last sum is going over all  ``energies'' (\ref {prop}) of the diagram lines, which are crossed by the dotted line, where $k_i$ are the momenta of the corresponding lines.

 Than  integrand in the momentum representation for the first diagram on the right-hand side (\ref{tv}) has the form:
\begin{equation}
\begin{matrix}
\includegraphics[angle=0, width=0.2 \textwidth]{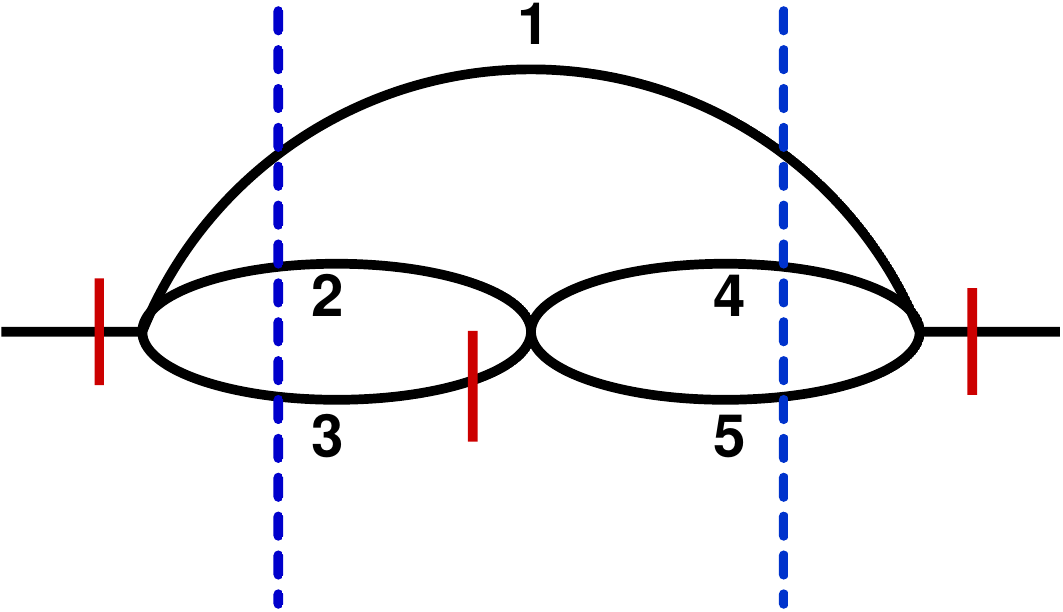}   \label{vremversii}
\end{matrix} \sim \frac{1}{E_1E_2E_4E_5} \cdot  \frac{1}{(E_1+E_2+E_3)} \cdot \frac{1}{(E_1+E_4+E_5)} \,.
\end{equation}
Here and in the following,  we denote $E_i \equiv E_{k_i}$. 
The integrands of the remaining time versions, shown in the figure (\ref{tv}), are constructed in a similar way.


\section{Reduction of diagrams}
\label{sec:reduction}
 A complicating circumstance in the problems of critical dynamics, in comparison with the static case, is a significantly larger number of momentum integrals arising as a result of integration over time (and a more complicated form of them). In a number of papers \cite{H model1, Peliti, H model, Folk1} the fact that, turning to certain sums of diagrams, one can appreciably simplify the integrands was used.
We propose a systematic procedure for such a reduction of diagrams, that makes it possible to automate the calculations, which is necessary for calculations in the higher orders of perturbation theory.

The possibility of such a reduction is actually seen from the relation (\ref{ZZ}). The equality (\ref{ZZ}) of static and dynamic counterterms is a consequence of a more general statement about the coincidence of 1-irreducible static functions $\langle \psi \psi \rangle_{1-irr}|_{st}$, $\langle \psi\psi\psi\psi\rangle_{1-irr}|_{st}$ and dynamic functions $\langle\psi'\psi \rangle_{1-irr}$, $\langle\psi'\psi\psi\psi \rangle_{1-irr}$  at zero frequency:
\begin{equation}\label{ZZZ}
\langle\psi'\psi \rangle_{1-irr}|_{\omega=0}=\langle \psi \psi \rangle_{1-irr}|_{st}\,, \quad \langle\psi'\psi\psi\psi \rangle_{1-irr}|_{\omega=0}=\langle \psi\psi\psi\psi\rangle_{1-irr}|_{st}\,.
\end{equation}
In the diagram language the equalities (\ref{ZZZ}) mean that for these functions the sum of the dynamic diagrams is reduced to a simpler object -- to the sum of the static diagrams. We will consider examples of the technical implementation of such a procedure, and then apply similar techniques to simplify the function of interest to us $\langle\psi'\psi' \rangle_{1-irr}|_{\omega=0}$.

Let us prove the equality:
\begin{eqnarray} \label{dyn=st}
\frac{1}{2} 
\begin{matrix}
\includegraphics[angle=0, width=0.13\textwidth]{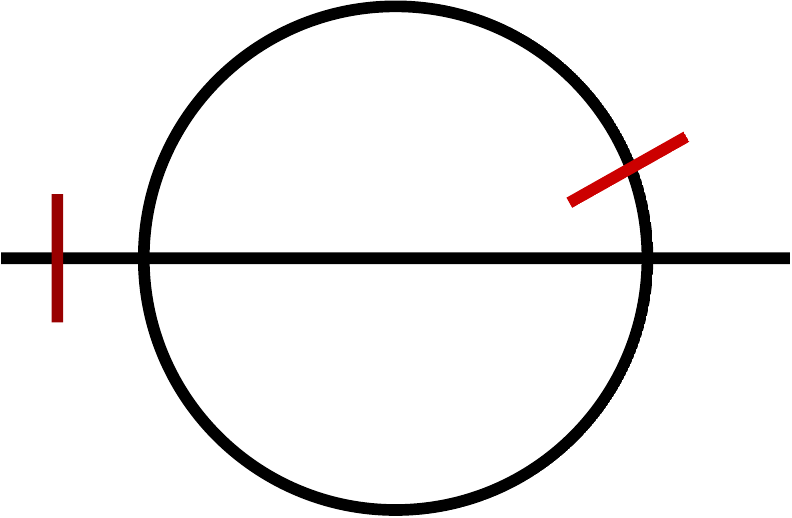}
\end{matrix} \mid_{\omega = 0} = 
\left[ \frac{1}{6}
\begin{matrix}
\includegraphics[angle=0, width=0.13\textwidth]{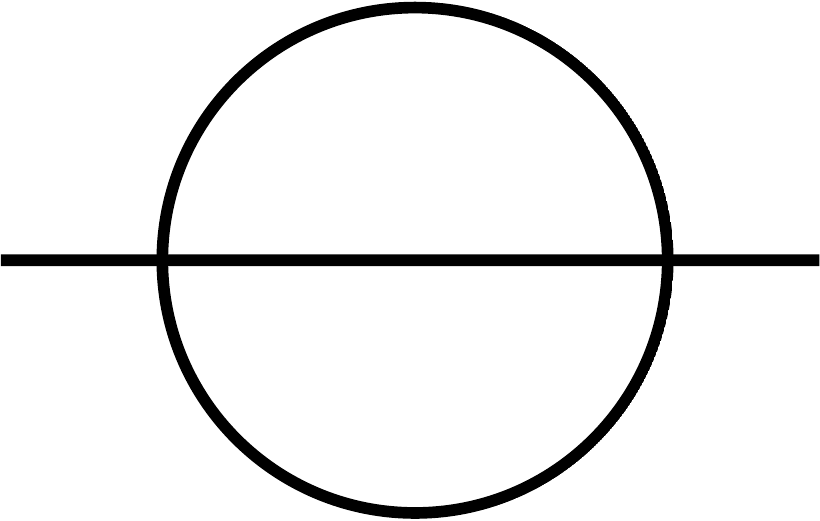}
\end{matrix}
\right]_{st}  \quad ,
\end{eqnarray}
where $\frac{1}{2}$ and $\frac{1}{6}$ are symmetry coefficients of diagrams.
Performing integration over time, we can write the left-hand side as
\begin{eqnarray} 
\frac{1}{2} 
\begin{matrix}
\includegraphics[angle=0, width=0.13\textwidth]{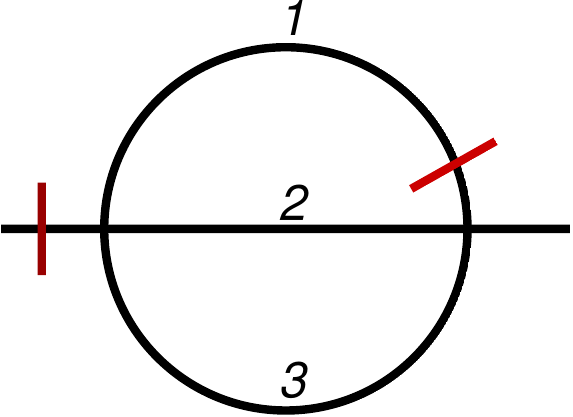}
\end{matrix} \mid_{\omega = 0} \quad 
 =  \quad 
 \frac{1}{2} \cdot
\begin{matrix}
\includegraphics[angle=0, width=0.13\textwidth]{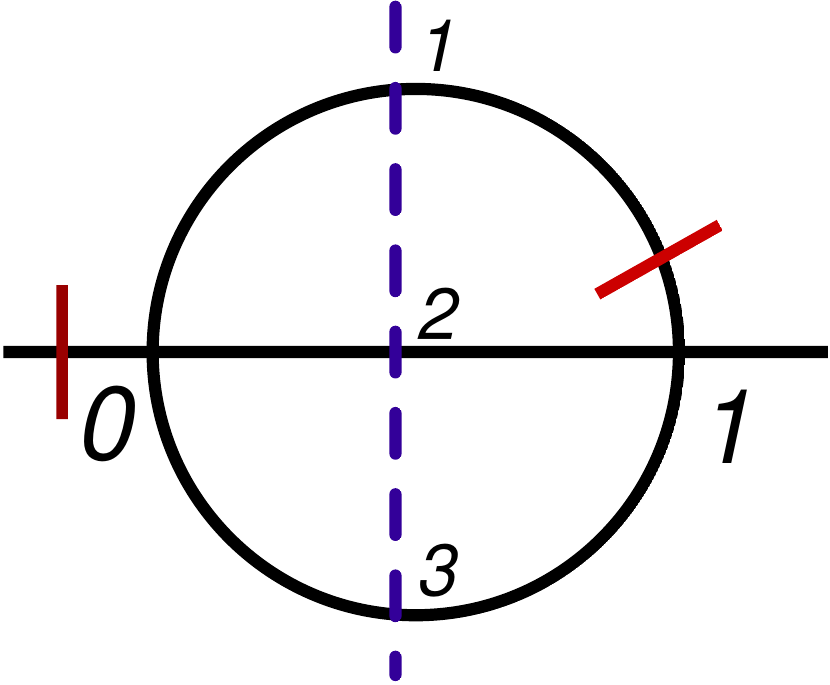}
\end{matrix}
\sim  \frac{1}{2} \frac{1}{E_2E_3} \cdot \frac{1}{(E_1+E_2+E_3)} 
\end{eqnarray}
Here the digits denote  the integration momenta, and the integrand stands on the right-hand side of the formula.
Symmetrizing this expression with respect to the momenta of integration, we obtain
\begin{eqnarray}
 &&\frac{1}{2} \frac{1}{E_2E_3}\cdot \frac{1}{(E_1+E_2+E_3)}  \rightarrow   \\
 && \rightarrow \frac{1}{6}  \left(\frac{1}{E_2E_3} +\frac{1}{E_1E_2}+\frac{1}{E_1E_3}\right)\cdot
  \frac{1}{(E_1+E_2+E_3)}=\frac{1}{6}  \frac{1}{ E_1 E_2 E_3}\quad ,  \nonumber
\end{eqnarray}
 which coincides with the integrand of the right hand side of (\ref{dyn=st}).\\
 In the diagram language, the symmetrization procedure can be written in the form:
 \begin{eqnarray}
\frac{1}{2} 
\begin{matrix}
\includegraphics[angle=0, width=0.10\textwidth]{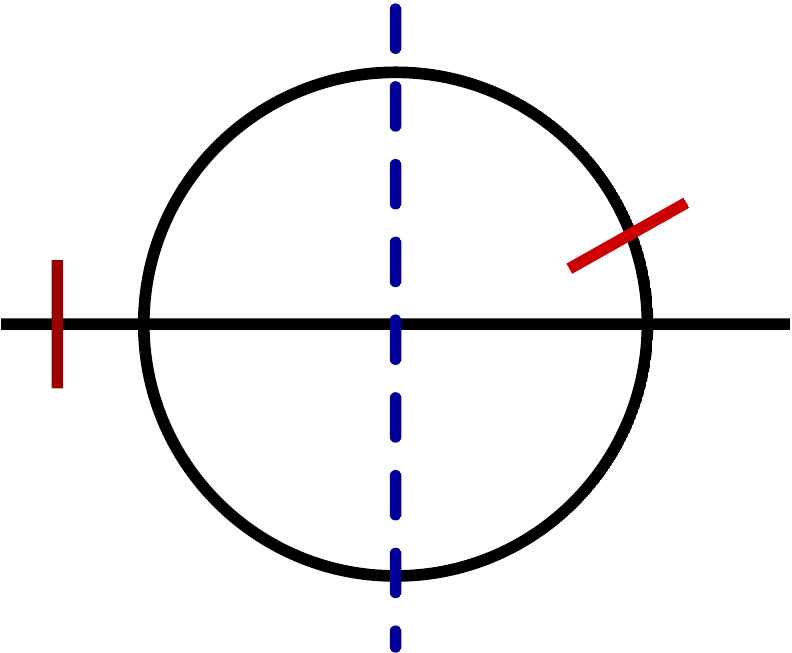}
\end{matrix} \quad 
 =   
   \frac{1}{6} \left( \begin{matrix}
\includegraphics[angle=0, width=0.10\textwidth]{arbuz-dyn-1-s1.pdf}
\end{matrix}+ 
 \begin{matrix}
\includegraphics[angle=0, width=0.10\textwidth]{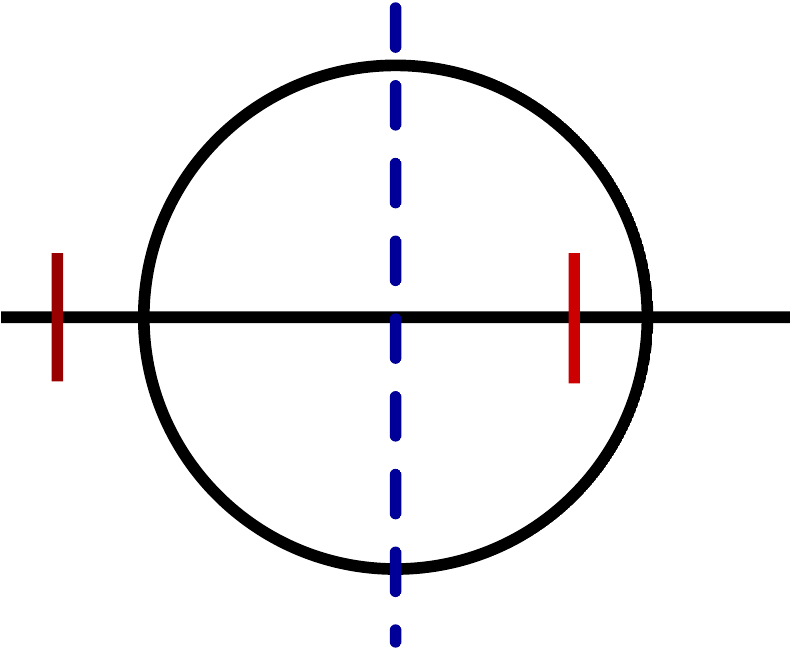}
\end{matrix}+ 
 \begin{matrix}
\includegraphics[angle=0, width=0.10\textwidth]{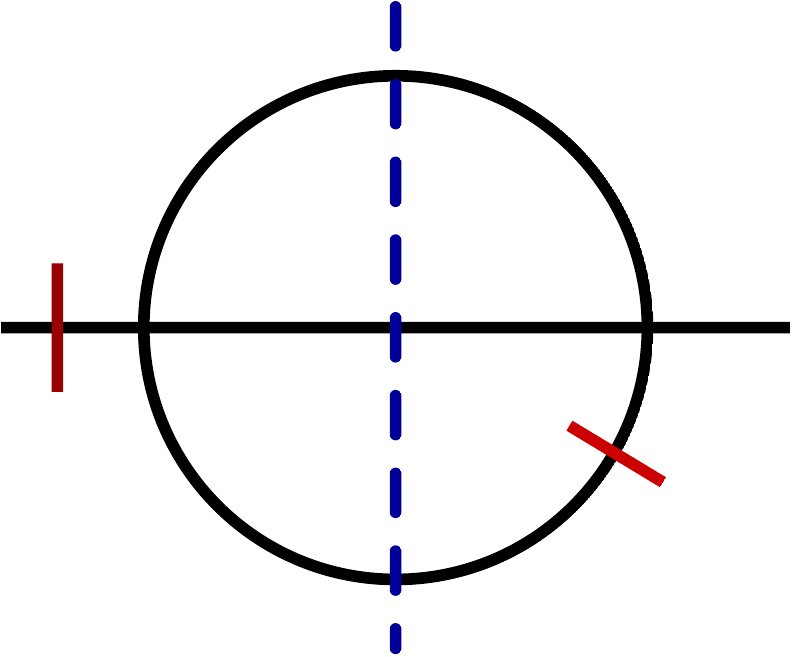}
\end{matrix}  \right)\,.
\end{eqnarray}
An analogous symmetrization for arbitrary diagrams can be written in the form of a symbolic equation: \\
\begin{eqnarray} \label{1hv}
 \begin{matrix}
\includegraphics[angle=0, width=0.5 \textwidth]{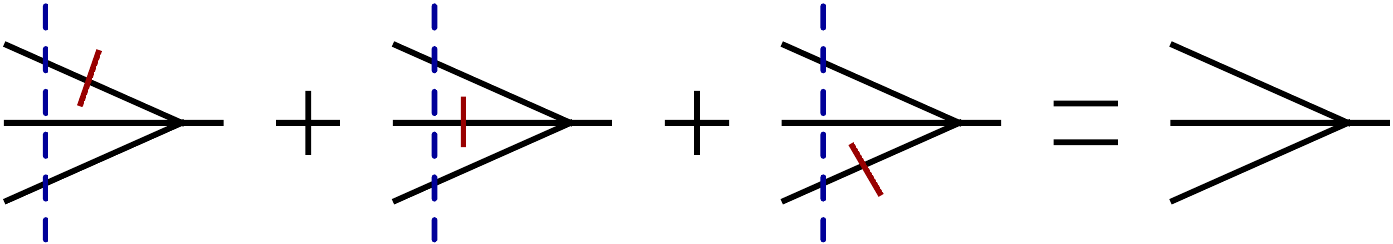}
\end{matrix}  
\end{eqnarray}
Obviously, the following two equations are also valid \\
\begin{eqnarray}
&& \begin{matrix}
\includegraphics[angle=0, width=0.4 \textwidth]{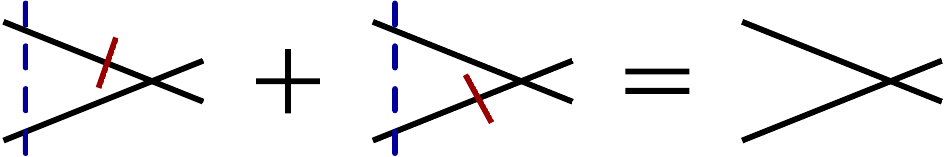}
\end{matrix}  \\
&& \begin{matrix} \label{0hv}
\includegraphics[angle=0, width=0.65 \textwidth]{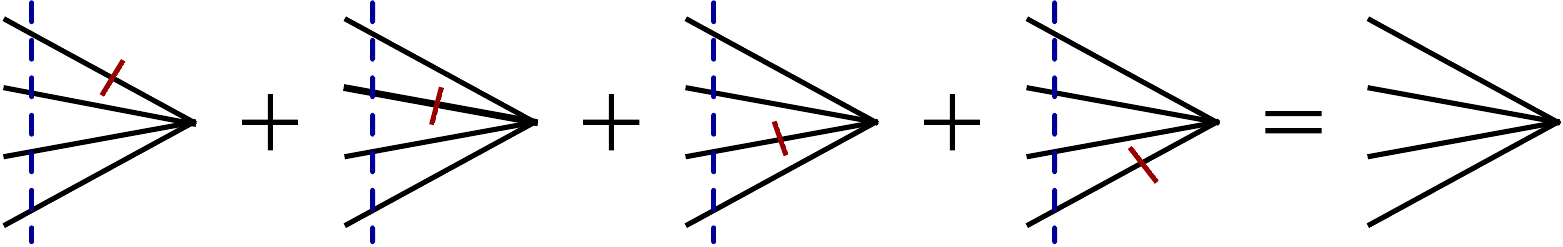}
\end{matrix}  
\end{eqnarray}
Using these equalities, we consider a more complicated example of the sum of three diagrams
\begin{equation}\label{eq1}
 J =\begin{matrix}
\includegraphics[angle=0, width=0.19\textwidth]{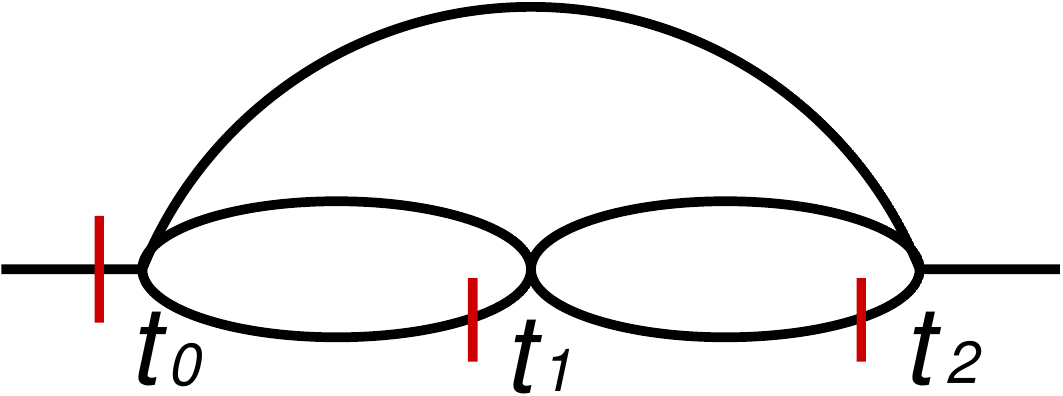}
\end{matrix} \mid_{\omega = 0}
+\frac{1}{2}
\begin{matrix}
\includegraphics[angle=0, width=0.19\textwidth]{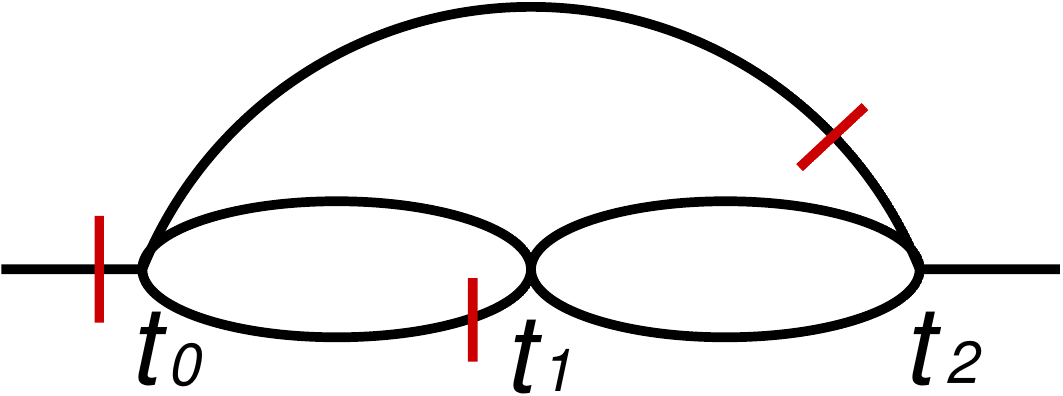}
\end{matrix} \mid_{\omega = 0}
+\frac{1}{2}
\begin{matrix}
\includegraphics[angle=0, width=0.19\textwidth]{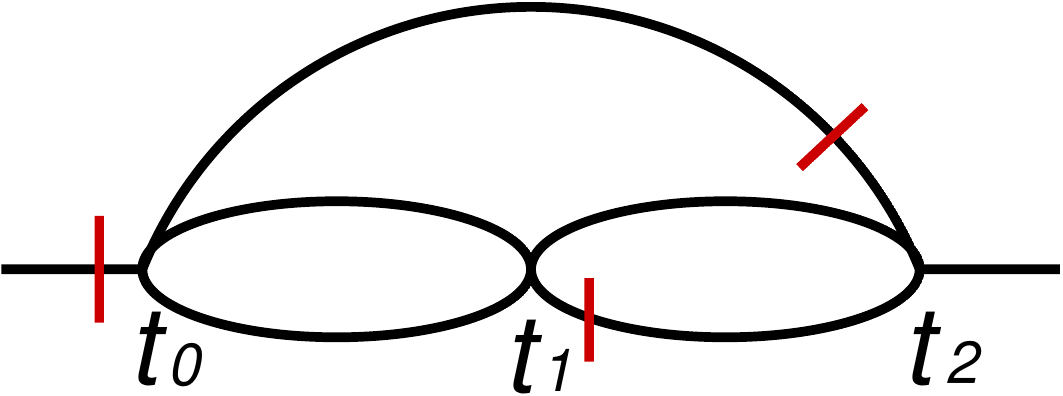}
\end{matrix} \mid_{\omega = 0}
\end{equation}
Calculating this sum using time versions and performing symmetrization,  we obtain:\\
\begin{equation}\label{eq2}
J = \frac{1}{2}\left(
\begin{matrix}
\includegraphics[angle=0, width=0.5\textwidth]{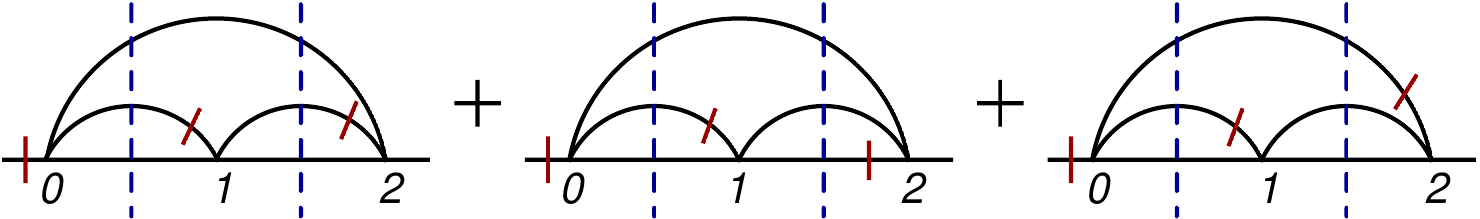}
\end{matrix}
\right)+ \frac{1}{2} \left(
\begin{matrix}
\includegraphics[angle=0, width=0.3\textwidth]{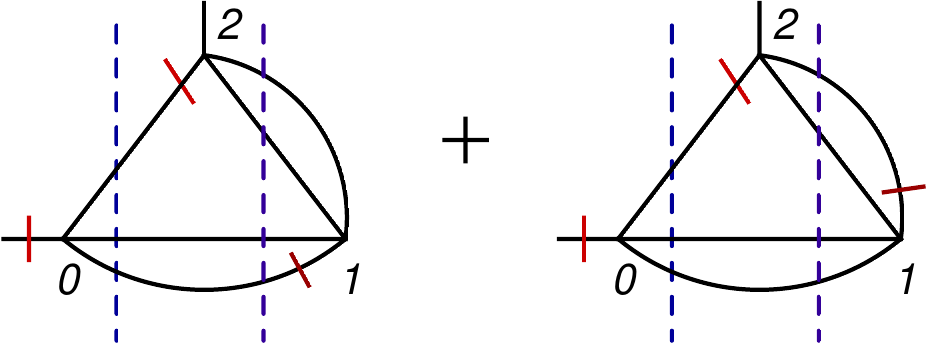}
\end{matrix}
\right)  
\end{equation}
The first diagram in (\ref{eq1}) has one time version and is divided into half the sum of the first two diagrams in (\ref{eq2}), the second diagram in (\ref{eq1}) has two time versions corresponding to diagrams 3 and 4 in (\ref{eq2}), and the last diagram has one time version corresponding to the fifth contribution in (\ref{eq2}). Using the relations (\ref{1hv})-(\ref{0hv}) in (\ref{eq2}), we find
\begin{eqnarray}
&&J =
\frac{1}{2}
\begin{matrix}
\includegraphics[angle=0, width=0.15\textwidth]{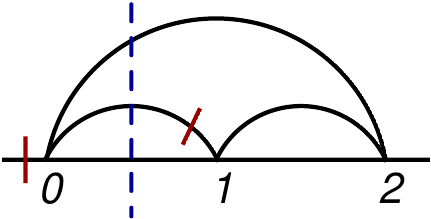}
\end{matrix} +
\frac{1}{4}
\begin{matrix}
\includegraphics[angle=0, width=0.11\textwidth]{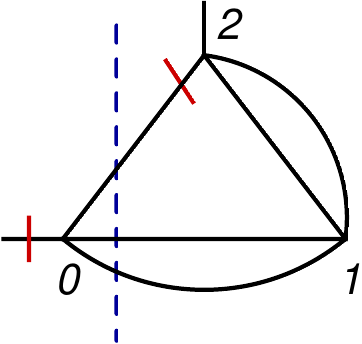}
\end{matrix} 
= 
\frac{1}{4}
\begin{matrix}
\includegraphics[angle=0, width=0.15\textwidth]{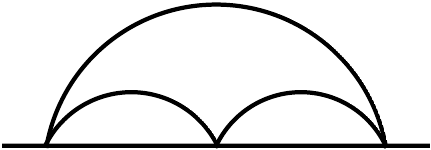}
\end{matrix} 
\end{eqnarray}
To simplify the first brackets in (\ref{eq2}) we used the equality (\ref{1hv}), for the second one -- (\ref{0hv}) and for the last transition -- (\ref{1hv}). As a result, the sum of the dynamic diagrams (\ref{eq1}) has been reduced to a single static one.
This example shows how  this technique reduces the number of diagrams.

As for diagrams of the one-irreducible function $\Gamma_{\psi'\psi'}$, a complete reduction to static diagrams is not possible. However, the use of the relations (\ref{1hv})-(\ref{0hv}) allows one even in this case to reduce significantly the number of contributions and to simplify their form. Let us consider  the sum of the last two diagrams in (\ref{tv}), which we rewrite as:
\begin{eqnarray}
J_1 = 
\begin{matrix}
\includegraphics[angle=0, width=0.11\textwidth]{sos-11-2-s.pdf}
\end{matrix}
+
\begin{matrix}
\includegraphics[angle=0, width=0.11\textwidth]{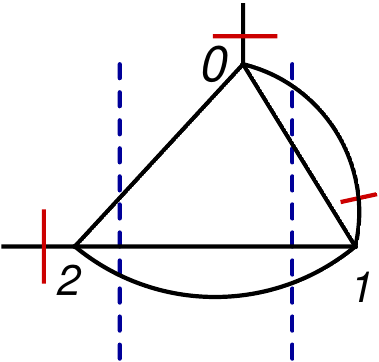}
\end{matrix}
\end{eqnarray}
 Taking into account that the values of the diagrams do not depend on the numbering of the vertices, making symmetrization and using the relation (\ref{0hv}), we obtain:
\begin{equation}
J_1 = \frac{1}{2} \left( 
\begin{matrix}
\includegraphics[angle=0, width=0.11\textwidth]{sos-11-2-s.pdf}
\end{matrix}
+\begin{matrix}
\includegraphics[angle=0, width=0.11\textwidth]{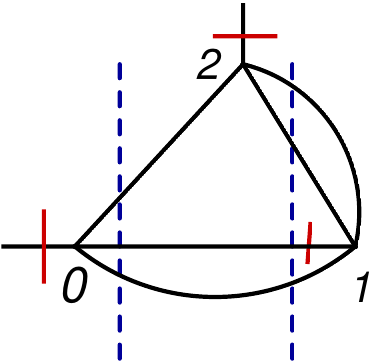}
\end{matrix}
+\begin{matrix}
\includegraphics[angle=0, width=0.11\textwidth]{sos-11-3-s-var1.pdf}
\end{matrix}+
\begin{matrix}
\includegraphics[angle=0, width=0.11\textwidth]{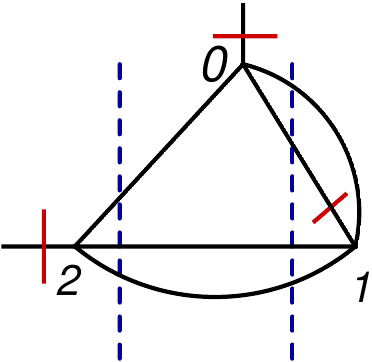}
\end{matrix}
\right) =\frac{1}{2}
\begin{matrix}
\includegraphics[angle=0, width=0.11\textwidth]{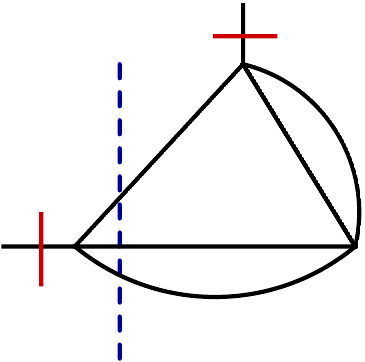}
\end{matrix}
\end{equation}
So one can see that even in this more complicated case reduction is possible.

Now let us formulate a general recipe for the reduction of diagrams,
 illustrating it with the example of the sum of the following two diagrams, containing in aggregate ten time versions:
\begin{eqnarray}
\begin{matrix}
\includegraphics[angle=0, width=0.5\textwidth]{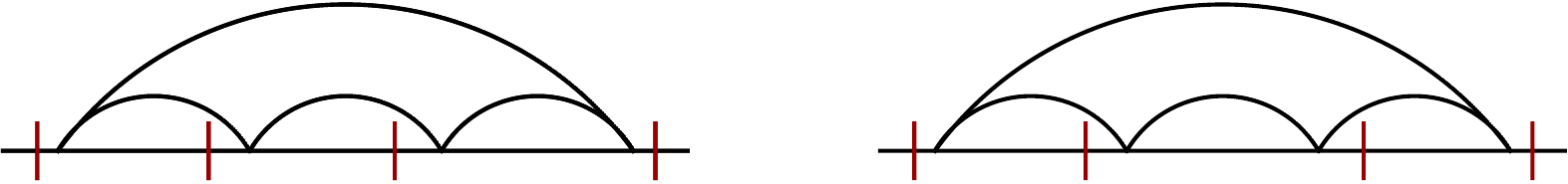}
\end{matrix} 
\end{eqnarray}
The result of the reduction is the sum of the diagrams constructed according to the following recipe.\\
\begin{itemize}
\item {\it Step I}. Draw the diagrams of the static theory so that vertices with external legs are extreme left and right, while other (internal) vertices ordered in all possible ways so that nearest vertices are connected to each other.\\
Possible order of vertices:
\begin{eqnarray}
\begin{matrix}
\includegraphics[angle=0, width=0.25\textwidth]{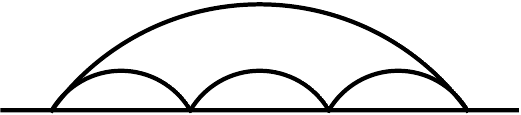}
\end{matrix} 
\end{eqnarray}
Forbidden one:
\begin{eqnarray}
\begin{matrix}
\includegraphics[angle=0, width=0.5\textwidth]{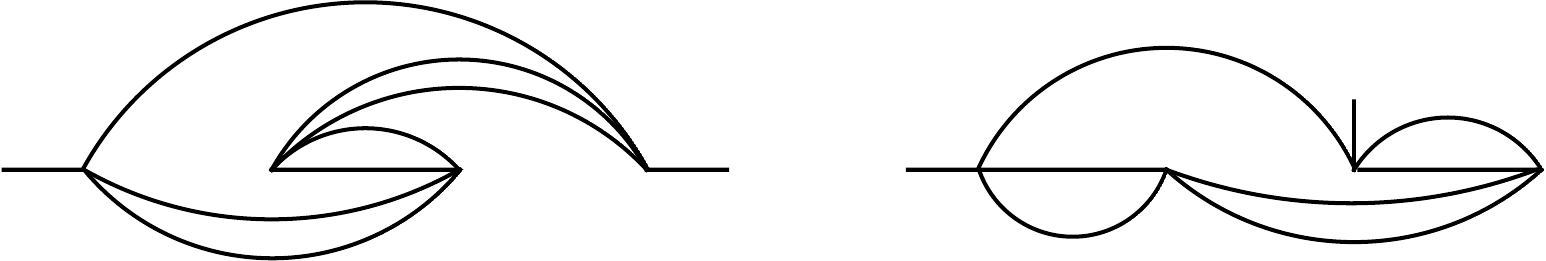}
\end{matrix}
\end{eqnarray}
\item {\it Step II}.
On basis of the diagrams from step I, draw a set of diagrams with dashed lines (from one to (number of vertices -1) sections) starting from the left vertex:
\\
\begin{eqnarray}
\begin{matrix}
\includegraphics[angle=0, width=0.63\textwidth]{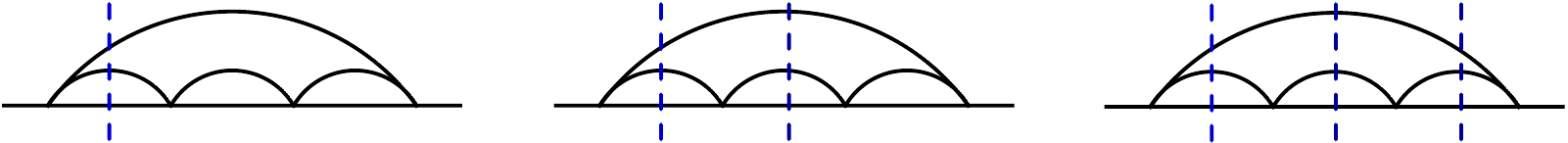}
\end{matrix}
\end{eqnarray}
\item {\it Step III}.
In all the lines coming from the left to the vertex located between the two sections, we arrange the strokes in all possible ways. If there are 2 similar lines, on which it is possible to arrange strokes, then we put only one, the remaining variant is taken into account by the symmetry coefficient.
\begin{eqnarray}\label{d1=3res}
\begin{matrix}
\includegraphics[angle=0, width=0.63\textwidth]{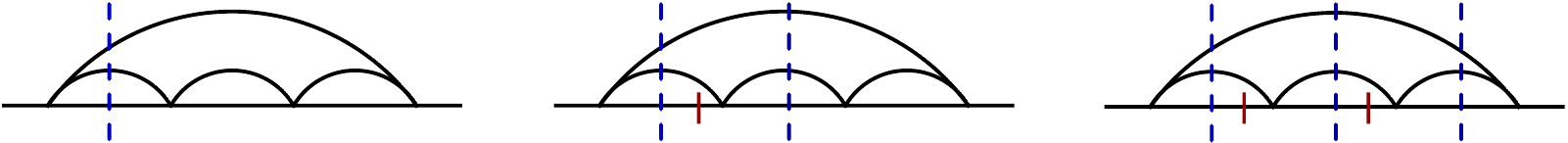}
\end{matrix}
\end{eqnarray}
\end{itemize}
 Thus, the original sum of ten time versions has been reduced to the sum of three effective diagrams (\ref{d1=3res}). (See another example  in the \ref{appendixa}) 
 
\section{Results}
\begin{figure}[h]
\centering
\subfloat[]{\includegraphics[angle=0, width=0.3\textwidth]{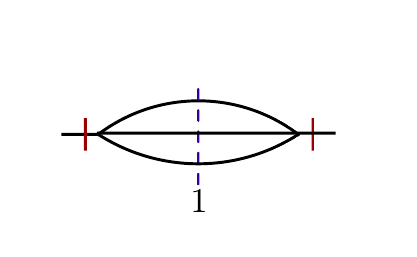}}
\subfloat[]{\includegraphics[angle=0, width=0.6\textwidth]{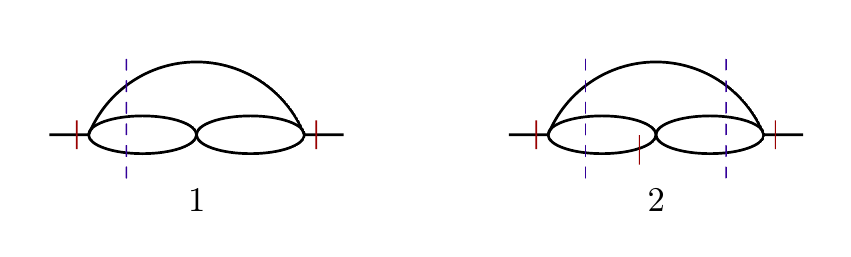}}
\caption{Diagrams of $\Gamma_{\psi'\psi'}$ after reduction: (a) two loops and  (b) three loops}
\label{fig:2loop}
\end{figure}

\begin{figure}[h!]
\centering 
\includegraphics[angle=0, width=1.0\textwidth]{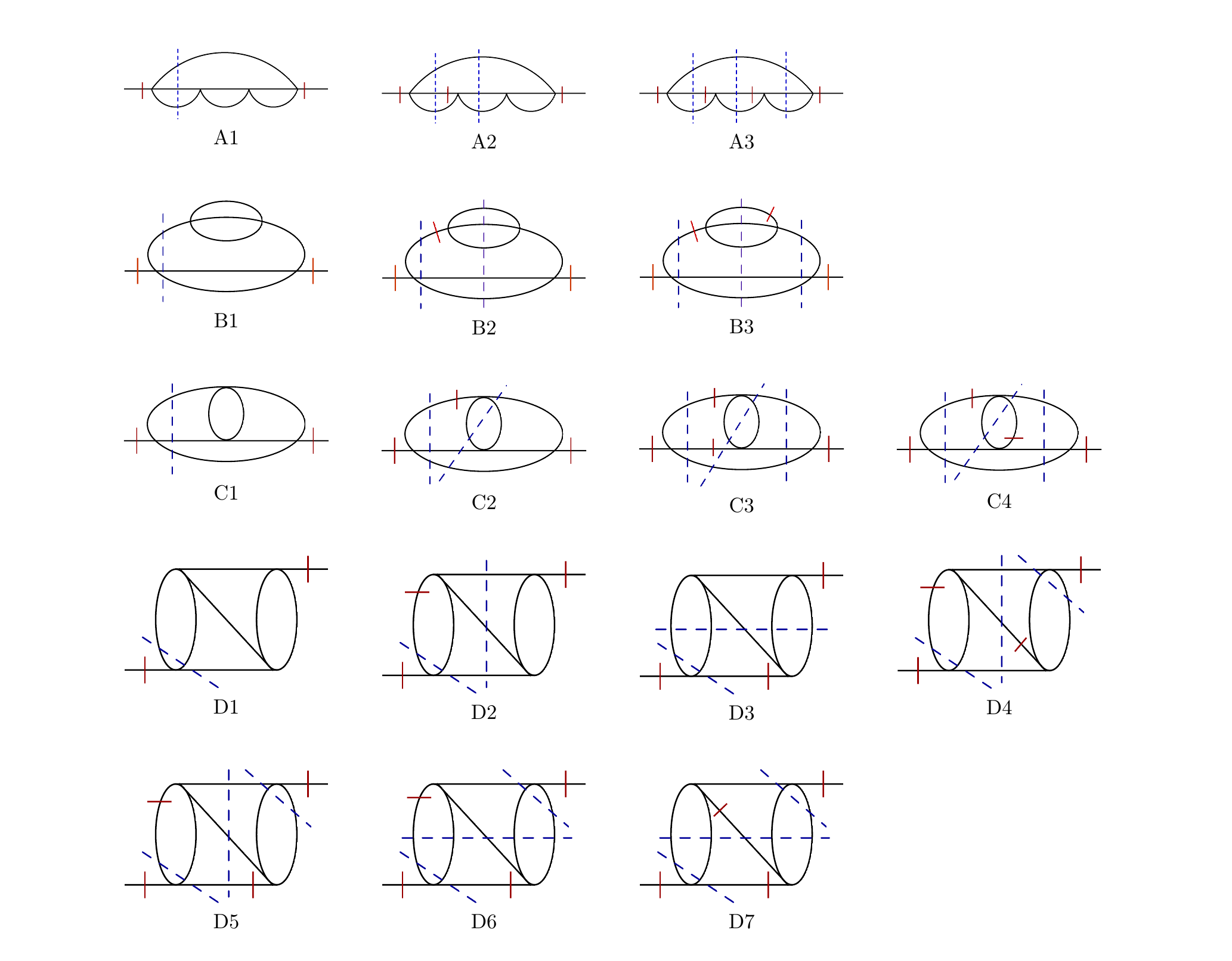}
\caption{Four loop diagrams of  $\Gamma_{\psi'\psi'}$ after reduction grouped by graph topology}
\label{fig:4loop}
\end{figure}

The result of the reduction of the diagrams of $\Gamma_{\psi'\psi'}$  up to the four-loop approximation is  depicted on Figs.~\ref{fig:2loop}-\ref{fig:4loop}.
The diagrams were calculated in the Feynman representation using the Sector Decomposition method~\cite{SD}. The required number of terms of their $\varepsilon$-expansion is given in the table~\ref{tab:res}, which also shows the corresponding symmetric factors $S$ and additional weight factors $f(n)$, which allow one to turn from the results for the one-component field with $n=1$ to the results for an $n$-component $O(n)$-symmetric model. The coefficients $A^{(m)}$ in the expansion (\ref{gamma}) are determined from the data of the table by the relation
\vspace{0.2cm}
\begin{eqnarray}\label{Gammam}
A^{(m)} = \sum_j S^{(m)}_j f^{(m)}_j(n) D^{(m)}_j(\varepsilon) \,.
\end{eqnarray}
Here
\begin{equation}
k_1=\frac{n+8}{9}\,, \quad k_2=\frac{n^2+6n+20}{27}\,, \quad k_3=\frac{n+2}{3}\,, \quad k_4=\frac{5n+22}{27}\,.
\end{equation}
\\
\\
 {
 \begin{table}
 \begin{tabular}[c]{|c|c|c|c|c|c|} \hline
\rule{0cm}{0.1cm}
\textnumero &$S^{(m)}$& \multicolumn{3}{|c|}{$D^{(m)}$ }&  $f^{(m)}(n)$ \\ \hline
\multicolumn{6}{|c|}{ $m = 2$ (2 loop)}     \\ \hline
& & $\varepsilon^{-1}$ & $\varepsilon^{0}$ &$\varepsilon^{1}$ & \\ \hline
1&$1/6$ &$ 0.2157615526(10) $&$  -0.3853514975(25) $ &$0.5692846610(35)$ &   $k_3$ \\  \hline
\multicolumn{6}{|c|}{  $m = 3$ (3 loop) }     \\ \hline
& & $\varepsilon^{-2}$ & $\varepsilon^{-1}$ &$\varepsilon^{0}$ & \\ \hline
1&$1/4$  & $ 0.143841039(8) $& $-0.330633628(30) $ & $  0.61007974(9) $ &    \multicolumn{1}{c|}{\multirow{2}{*}{$k_1\,k_3$}}  \\  \cline{1-5}
2& $1/2$&  $ 0.071920514(4) $&$-0.196352188(16) $& $  0.41485450(4) $ &   \\  \hline
\multicolumn{6}{|c|}{$m = 4$ (4 loop)}     \\ \hline
& & $\varepsilon^{-3}$ & $\varepsilon^{-2}$ &$\varepsilon^{-1}$ & \\ \hline 
A1& $1/8$ & $0.10788085(9)$ & $-0.3032746(4) $ & $ 0.6479956(16) $ & \multicolumn{1}{c|}{\multirow{3}{*}{$k_2\,k_3$}}  \\  \cline{1-5}
A2& $1/4$ & $0.05394036(5) $ & $ -0.17491376(20)$ & $ 0.4258852(8) $ & \\  \cline{1-5}
A3& $1/2$ & $ 0.026970186(22) $ & $-0.09909542(10)$ & $ 0.2689090(4) $ & \\ \hline
B1& $1/12$ &  & $ -0.1004829(4) $ & $ 0.1431086(18)$ &   \multicolumn{1}{c|}{\multirow{3}{*}{$k_3^2$}}  \\  \cline{1-5}
B2& $1/12$ & & $ 0.023276508(28)$ & $ -0.08913734(14)$ & \\  \cline{1-5}
B3& $1/4$ &  & $ -0.01363165(4) $ & $ 0.03119582(18) $ & \\ \hline
C1& $1/4$ & $0.05394039(4) $ & $ -0.12466740(20) $ & $ 0.2359704(10)$ &   \multicolumn{1}{c|}{\multirow{4}{*}{$k_3\,k_4$}}  \\  \cline{1-5}
C2& $1/2$ & & & $ 0.04114906(8) $ & \\ \cline{1-5}
C3& $1/2$ & & $0.011638248(13) $ & $ -0.05352246(7) $ & \\ \cline{1-5}
C4& $1$ & $0.013485088(11) $ & $ -0.04862411(5) $ & $ 0.12779878(22)$ & \\ \hline
D1& $1/4$ & $0.053940377(5) $ & $-0.124667192(32) $ & $0.28863410(18) $ &   \multicolumn{1}{c|}{\multirow{7}{*}{$k_3\,k_4$}}\\ \cline{1-5}
D2& $1/2$ & $0.05394044(5) $ & $-0.17491404(20) $ & $ 0.4107878(8) $ & \\   \cline{1-5}
D3& $1/4$ & & & $ 0.03941088(8) $ & \\ \cline{1-5}
D4& $1/2$ & $ 0.013485098(11) $ & $-0.04862416(5) $ & $ 0.14655896(22) $ & \\    \cline{1-5}
D5& $1/2$ & $0.0134850952(14) $ & $-0.048624182(8) $ & $ 0.13962442(4) $ & \\ \cline{1-5}
D6& $1/2$ & & $0.007758853(10)$ & $ -0.03059186(5) $ & \\ \cline{1-5}
D7& $1/4$ & & $0.0077588424(16)$ & $ -0.027130478(10) $ & \\ \hline
\end{tabular}
\caption{Values of $\varepsilon$-expansion of diagrams from Figs.~\ref{fig:2loop}-\ref{fig:4loop}}
\label{tab:res}
\end{table}
}
Renormalization constant $Z_g$ in $(\ref{gamma})$ is known from the statics and with the required  accuracy is equal to 
\begin{eqnarray}\label{Zg}
Z_g&=&1 + u \frac{8+n}{6\varepsilon}+u^2 \left(\frac{(8 + n)^2}{36 \varepsilon^2}-\frac{14+3n}{24\varepsilon} \right)+O(u^3)\,.
\end{eqnarray}
Regarding the value of ${Z_\tau}$ in $(\ref{gamma})$, we need to make the following remark. The use of the static renormalization constants $Z_g$ and ${Z_\tau}$ in $(\ref {gamma})$ implies that in calculating of the function $\Gamma_{\psi'\psi'}$  one takes into account all diagrams, including ones that contain tadpole subgraphs. It is known that such diagrams can be ignored (which we did), if we do not take into account the tadpoles in counterterms as well. So, if tadpoles are not taken into account while calculating $\Gamma_{\psi'\psi'}$, than to be consistent we should remove corresponding contributions from  the renormalization constant ${Z_\tau}$.
The resulting renormalization constant $\tilde{Z_\tau}$ with the necessary accuracy is given by the expression 

\begin{eqnarray}\label{Ztau}
\tilde{Z_\tau} &=&1-u^2\left(\frac{2+n}{12 \varepsilon^2} + \frac{5 (2 + n)}{144 \varepsilon}\right)+O(u^3) \,.
\end{eqnarray}
Note, that the problem of tadpoles is absent if the calculations of the renormalization constants are carried out in the ``massless''  theory with $\tau=0$, in which the tadpoles are defined by zeros. In this theory, instead of $(\ref{gamma})$, the value $\Gamma_{\psi'\psi'}|_{\omega=0, \tau=0} $ is calculated for which the factor $ (\mu^2/(\tau Z_{\tau}))^{\varepsilon/2}$ on the right side of $(\ref{gamma})$ is replaced by $(\mu/p)^{\varepsilon}$. However, with this approach, the integrands in the Feynman representation are slightly more complicated.

Substituting the expressions (\ref{Z1}), (\ref{Zg}), (\ref{Ztau}) in (\ref{gamma}), and calculating  values of $A^{(m)}$ with   (\ref{Gammam}), we can find the coefficients $z_{nm}$ from the requirement of the cancellation of the pole contributions in $\varepsilon$. According to renormalization theory, the coefficients $z_{nm}$ at the highest poles in $\varepsilon$ ($m>1$) are expressed in a certain way in terms of the coefficients $z_{n1}$ of the first poles, which guarantees a cancellation of pole contributions  in (\ref{gamma}). This fact can be used as an additional self-consistency check for the multi-loop renormalization group calculations.

The renormalization constant $Z_1$ is associated with the RG-function $\gamma_1$
\begin{equation}\label{gamma1}
\gamma_1(u)=\beta(u)\,\partial_u \log Z_1.
\end{equation}
The expression for the $\beta$-function is currently known with six-loop accuracy~\cite{komp6,mkomp6}. We do not need its explicit form, since the connection mentioned above between the coefficients at the higher poles with the coefficient at the first pole makes it possible to represent $\gamma_1(u)$ in a simpler form
\begin{equation}\label{gamma1a}
\gamma_1(u)=- u\partial_u
(z_{21}u^2+z_{31}u^3+z_{41}u^4+...).
\end{equation}

The dynamic critical exponent $z$ is expressed in terms of the value $\gamma_1^* \equiv \gamma_1 (u_*)$ of the function $\gamma_1(u)$ at the fixed point $u_*$ and the Fisher exponent $\eta$ by the relation \cite{Vasiliev}
\begin{equation}\label{z}
z=2+\gamma_1^*-\eta\,.
\end{equation}
Substituting the values of $z_{n1}$ into (\ref{gamma1a}) and normalizing the result to the value of the two-loop contribution, we obtain:
 \begin{equation}\label{gam}
 \gamma_1^*=k_3 h \frac{u_*^2}{24}\left[1+b_1k_1u_*+(b_2k_2+b_3k_3+b_4k_4)u_*^2 \right]+O(u_*^5)\,,
    \end{equation}
   where
\begin{eqnarray}\label{b}
&& h = 6 \ln(4/3)\simeq 1.726092433\,,\\ 
&& b_1= -0.4939306(5)\,,   \quad b_2= -0.251043(19)\,,\\ 
&& b_3= -0.169990(9)\,,  \quad b_4= 1.806593(30)\,.  
\end{eqnarray}
    The value $u_*$, determined by the condition $\beta(u_*)=0$, with the required accuracy is given by the expression \\
    \begin{eqnarray}\label{uzv}    
u_* &=& \frac{6}{n+8} \varepsilon + \frac{18 (3 n+14)}{(n+8)^3}\varepsilon^2+ 
+\frac{3}{4(n+8)^5}\Big(-33 n^3+ 110 n^2+ 1760 n+ \nonumber
\\
&& + 4544    -  96 (n+8) (5 n+22) \zeta(3)\Big)\varepsilon^3+O(\varepsilon^4) \,. 
\end{eqnarray}
The results of the dynamic exponent are usually presented in the form 
 \begin{equation}\label{zz}
 z=2+R \eta\,.
    \end{equation}
The value of $\eta$ can be written in a form similar to (\ref{gamma1a}), with the required accuracy
   \begin{equation}\label{etazv}
 \eta=k_3\frac{u_*^2}{24}\left[1+a_1k_1u_*+(a_2k_2+a_3k_3+a_4k_4)u_*^2 \right]+O(u_*^5)\,,
    \end{equation}
  where  
\begin{equation}\label{ai}
a_1=-\frac{3}{8}\,, \qquad  a_2=-\frac{15}{64}\,, \qquad  a_3=-\frac{5}{32}\,, \qquad  a_4=\frac{45}{32}\,.
\end{equation}
        From (\ref{zz}) and (\ref{z})  taking into account (\ref{gamma1a}), (\ref{etazv}), (\ref{uzv}) we get 
\begin{equation}\label{RR}
 R=\left(6\,ln(4/3)-1\right)\left[1+c_1\varepsilon+\left(c_2+\frac{(c_3+c_4n)}{(n+8)^2}\right)\varepsilon^2+{O}(\varepsilon^4)\right]\,,
\end{equation}
where the coefficients $c_i$ are determined by the relations
\begin{eqnarray}\label{c1}
&& c_1= \frac{2}{3} \frac{h}{h-1}(b_1- a_1)\,,\nonumber \\
&& c_2=\frac{4h}{3(h-1)}\left( \frac{1}{3}a_1(a_1-b_1) +(b_2-a_2)\right)\,,\\
&& c_3=\frac{4h}{3(h-1)} \left( 21 (b_1-a_1)- 44(b_2-a_2) +18(b_3-a_3)+22(b_4-a_4)\right)\,,\nonumber \\ 
&& c_4=\frac{2h}{3(h-1)} \left( 9 (b_1-a_1)- 20(b_2-a_2) +18(b_3-a_3)+10(b_4-a_4)\right)\,.\nonumber
\end{eqnarray}
The first two terms of the $\varepsilon$-expansion (\ref{RR}) do not depend on the number of components of the field $n$. The first of them was calculated in the work \cite{1}, the second -- in the work \cite{2}, where  the expression for $b_1$ was obtained
\begin{equation}\label{b1}
 b_1=\frac{\pi^2/8-F(1/4)}{\ln(4/3)}-\frac{3}{4}+\frac{13}{8} \ln4-\frac{21}{8}\ln3\simeq-0.493930232\,,
    \end{equation}
    \begin{equation}\label{F}
 F(x)=\int_x^1 \frac{\ln t}{t-1}dt\,,
    \end{equation}
which, according to (\ref{ai}), (\ref {c1}) and (\ref{b1}) corresponds to
\begin{equation}\label{c1a}
c_1 \sim -0.188483416\,.
    \end{equation}
    
In the work
\cite{1} the value $R$ was calculated in the leading order of the $1/n$ expansion for an arbitrary dimension $d$:
    \begin{equation}\label{Rinf}
    R_{\infty}=\frac{4}{4-d}\left(\frac{d\Gamma^2(d/2-1)/\Gamma(d-2)}{8\int_0^{1/2}dx[x(2-x)]^{d/2-2}}-1\right)\,.
    \end{equation}
    The first terms in the expansion of this quantity with respect to $\varepsilon=4-d$ have the form
    \begin{equation}\label{Rinf3}
    R_{\infty}=\left(6\,ln(4/3)-1\right)\left(1-0.188483417\,\varepsilon-0.099952926\,\varepsilon^2+O(\varepsilon^3)\right)\,.
    \end{equation}
        Taking into account that the first two terms of the $\varepsilon$-expansion (\ref{RR}) do not depend on $n$, they coincide with the corresponding contributions to (\ref{Rinf3}), which is confirmed by the results of \cite{1}, \cite{2}. The expansion (\ref{Rinf3}) also determines the coefficient $c_2$ in the quadratic by $\varepsilon$ contribution in (\ref{RR}):
     \begin{equation}\label{c2}
   c_2=-0.099952926\,.
    \end{equation}   
    
    The values of the coefficients $c_i$ in (\ref{RR}) obtained in this paper are 
    \begin{equation}
    \begin{split}
 c_1 = -0.1884840(7) ,& \qquad c_2 = -0.09995(6),\\ 
 c_3 = 21.5412(34) ,& \qquad c_4 = 4.7847(8) .
\end{split}
\label{results_ci}
\end{equation}
Results obtained are in full agreement with three loop calculations \cite{2}, as well as with $1/n$-expansion~\cite{1}, as for the four loop contribution, the coefficients $c_i$ were first calculated in \cite{Sladkoff} by a different method with much less accuracy:
 \begin{equation}
 c_1=-0.1884(9),\,\,c_2=-0.100(4),\,\,c_3=21.5(4), \,\,c_4=4.78(6)\,,
 \end{equation}
as one can see our results are in agreement with this calculations as well.

Here we also give the resulting $\varepsilon$-expansion directly for the dynamic index $z$ for $n=1$:
\begin{equation}
z = 2+ 0.0134461561 \varepsilon^2 +0.011036273(10)\varepsilon^3  -0.0055791(5) \varepsilon^4+O(\varepsilon^5)\,.
\end{equation}

    \section{Conclusion}
        
In this paper we performed four loop calculation of the critical exponent $z$ in the framework of $\varepsilon$-expansion and renormalization group. To perform this calculation we developed a method of reduction of the diagrams in the models of critical dynamics which allows to significantly reduce a number of diagrams to be calculated. This method combined with Feynman representation and Sector Decomposition technique~\cite{SD} allows us to obtain high precision numbers for four-loop contribution to dynamic exponent.

The necessity of high loop calculations for model A was pointed out in 
\cite{Nalimov} where Borel resummation of the results of the work~\cite{Sladkoff} was performed. It was shown that results of resummation are very  sensitive to particular realizations of the summation, which must be a consequence  of the insufficient number of terms of the $\varepsilon$-expansion.
As it was noted, the model A is the simplest model of critical dynamics, for more complicated models renormalization group calculations are limited at maximum by two loop order. The lack of perturbative information in this models does not allow to make solid theoretical predictions, moreover in some models (e.g. model E) it is not possible to confidently distinguish concurrent asymptotic regimes.

The method discussed in this paper allows to significantly reduce the total calculation time for such problems and opens the possibility to extend this calculations to higher loops and more complicated models. For example, in model A at 5 loop level number of diagrams is reduced from 1025 to 201 and with more simple integrands, which gives us a possibility to reach high accuracy of numerical calculations. While for more complicated theories like model E of critical dynamics \cite{obzor1,Emodel,Peliti,dohm} (where RG calculations are limited only by 2 loop order) our preliminary estimations show that this factor may be even greater than 5 and  this gives us a hope that 3 and 4 loop calculations in this models can be feasible.



\section*{Acknowledgements}
Research was carried out using computational resources provided by Resource Center  ``Computer Center of SPb'' (http://www.cc.spbu.ru/en).\\
Figures of Feynman graphs in this article were
created with JaxoDraw \cite{jaxodraw}

\appendix   
    \section{Example of diagram reduction}
    \label{appendixa}
As a second example of the diagram reduction, let us consider the sum of 28 time versions of the diagrams of the $\Gamma_{\psi'\psi'}$ of the following type:
\begin{equation}\label{diagr2}
\begin{matrix}
\includegraphics[angle=0, width=0.19\textwidth]{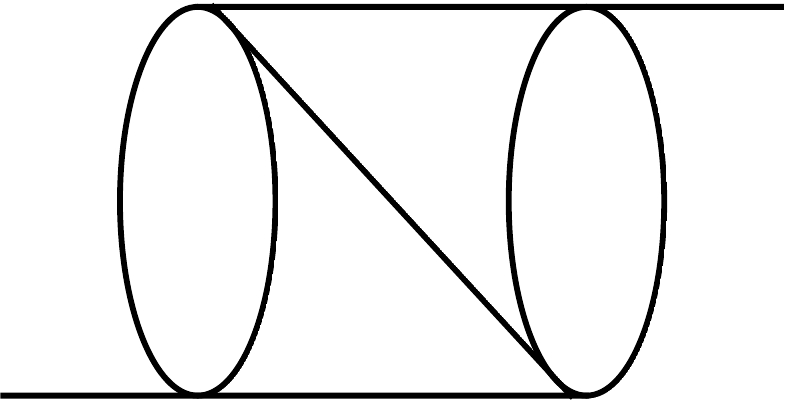}
\end{matrix} 
\end{equation}
\begin{itemize}
\item {\it Step I}.
\begin{eqnarray}
\begin{matrix}
\includegraphics[angle=0, width=0.5\textwidth]{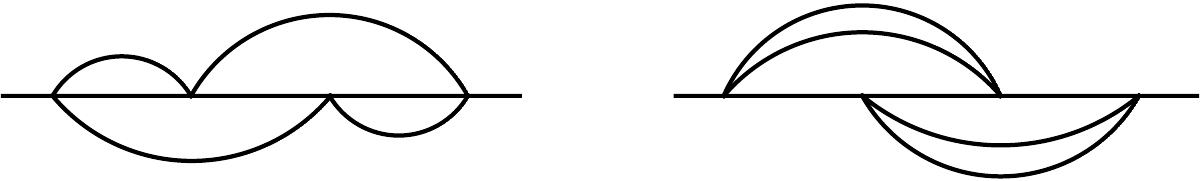}
\end{matrix}  
\end{eqnarray}
\item {\it Step II}. 
\begin{eqnarray}
\includegraphics[angle=0, width=0.65\textwidth]{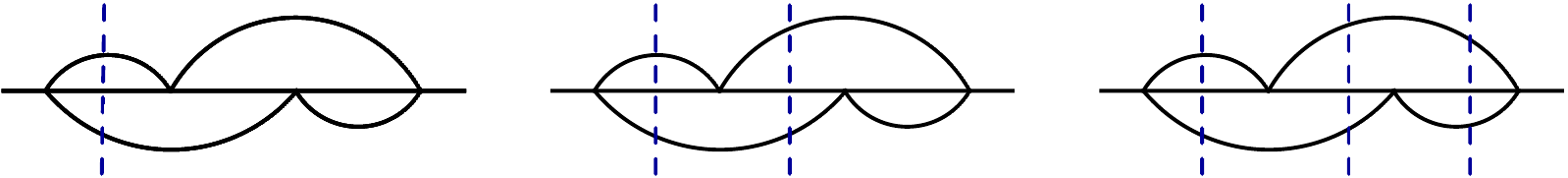}
  \nonumber\\
\includegraphics[angle=0, width=0.7\textwidth]{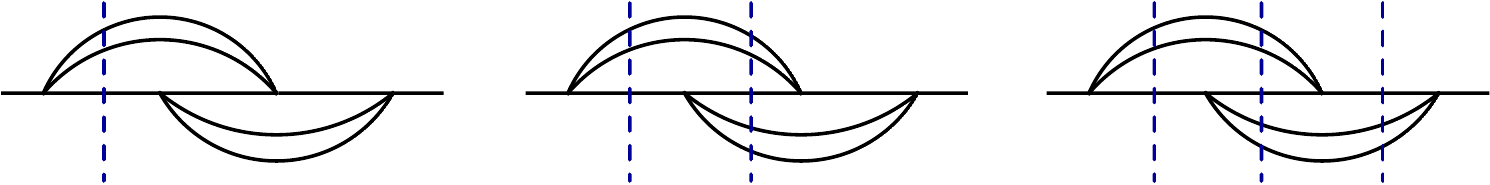}
\end{eqnarray}
\item {\it Step III}. 
\begin{eqnarray}
\includegraphics[angle=0, width=0.7\textwidth]{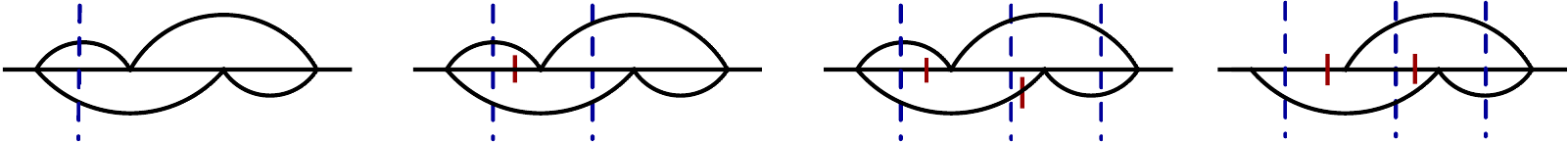}  \nonumber \\
\includegraphics[angle=0, width=0.85\textwidth]{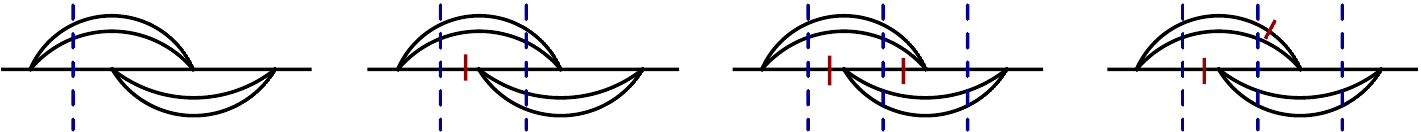}
\label{d2=8res}
\end{eqnarray}
\end{itemize}
   As a result, 28 versions of the diagrams (\ref{diagr2}) were reduced to 8 diagrams (\ref{d2=8res}). 
    
    \section{Feynman presentation}   
    \label{appendixb}
    The dependence on the integration momenta in the diagrams after the integration over time has a structure that makes it possible to turn to the Feynman representation.
     This allows us to use the Sector Decomposition method \cite{SD}, as in problems of critical statics \cite{5loop}.
     As an example we will consider the second diagram from (\ref{d1=3res}):
   \begin{eqnarray}\label{dyn2}
\begin{matrix}
\includegraphics[angle=0, width=0.35\textwidth]{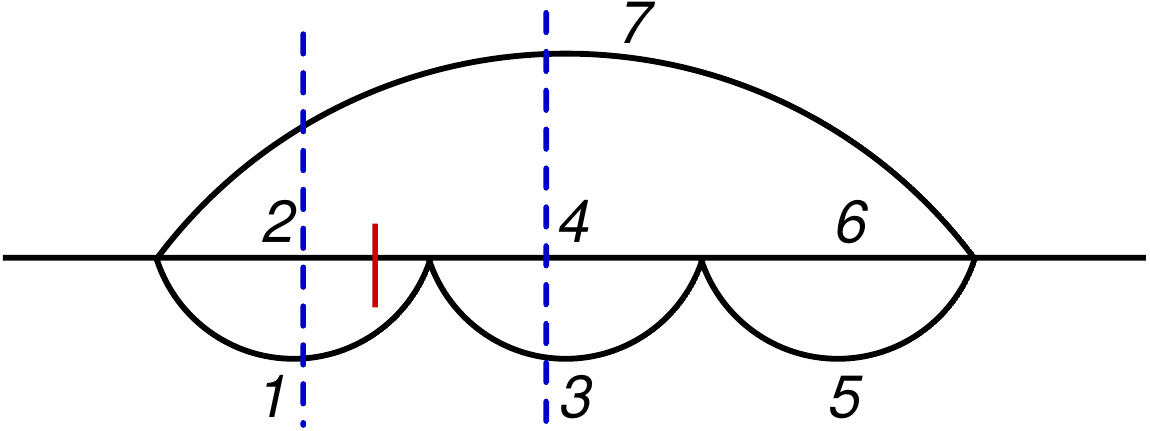}
\end{matrix} 
\end{eqnarray}
The numbers on the lines denote the integration momenta flowing from left to right. 
The integral that corresponds to diagram \eqref{dyn2} looks as follows:
\begin{equation}\label{J}
J=\int d{\bf k}_1 ... \int d{\bf k}_7\frac{\delta({\bf k}_1+{\bf k}_2+{\bf k}_7)\,\delta({\bf k}_5 +{\bf k}_6+{\bf k}_7)\,\delta({\bf k}_1 +{\bf k}_2-{\bf k}_3-{\bf k}_4)}{E_{k_1}E_{k_3}E_{k_4}E_{k_5}E_{k_6}E_{k_7}(E_{k_1}+E_{k_2}+E_{k_7})\,(E_{k_3}+E_{k_4}+E_{k_7})}\,.
\end{equation}
Associating each of the 8 factors in the denominator of  (\ref {J}) to the Feynman parameter $v_i$ and using the Feynman formula, we obtain:
\begin{equation}\label{J1}
J=\int_0^1 \prod d v_i\,\delta \left (\sum v_i -1\right)\,F(\{v\})\,,
\end{equation}
where
\begin{equation}\label{F}
F(\{v\})= \int d{\bf k}_1 ... \int  d{\bf k}_7\,\frac{\delta({\bf k}_1+{\bf k}_2+{\bf k}_7)\,\delta({\bf k}_5 +{\bf k}_6+{\bf k}_7)\,\delta({\bf k}_1 +{\bf k}_2-{\bf k}_3-{\bf k}_4)}{Q^\alpha}\,,
\end{equation}
\begin{eqnarray} \label{Q}
 Q=&&v_1E_{k_1}+v_3E_{k_3}+v_4E_{k_4}+v_5E_{k_5}+v_6E_{k_6}+v_7E_{k_7}+ \nonumber \\ 
 &&+v_8(E_{k_1}+E_{k_2}+E_{k_7})+v_9(E_{k_3}+E_{k_4}+E_{k_7}),
\end{eqnarray}
 $\alpha = 8$ -- the number of factors in the denominator of  (\ref{J}). Writing $Q$ in the form
 \begin{eqnarray} \label{Qu}
Q=u_1E_{k_1}+u_2E_{k_2}+u_3E_{k_3}+u_4E_{k_4}+u_5E_{k_5}+u_6E_{k_6}+u_7E_{k_7}\,,
\end{eqnarray}
where
 \begin{eqnarray}\label{uv}
u_1=v_1+v_8\,,\,\,u_2=v_8\,,\,\,u_3=v_3+v_9\,,\,\, u_4=v_4+v_9\,,\nonumber \\ u_5=v_5\,,\,\,u_6=v_6\,,\,\, u_7=v_7+v_8+v_9,
\end{eqnarray}
from (\ref{F}), (\ref{Q}) we obtain
\begin{equation}\label{F1}
F(\{v\})= \int d{\bf k}_1 ... d{\bf k}_7\,\frac{\delta({\bf k}_1+{\bf k}_2+{\bf k}_7)\,\delta({\bf k}_5 +{\bf k}_6+{\bf k}_7)\,\delta({\bf k}_1 +{\bf k}_2-{\bf k}_3-{\bf k}_4)}{(\sum_{j=1}^7 u_j E_{k_j} )^\alpha}\,.
\end{equation}
 Choosing in (\ref{F1}) as independent variables a certain set $\{{\bf k}_{i_1}, {\bf k}_{i_2}, {\bf k}_{i_3}, {\bf k}_{i_4} \}$, and performing the integration with the help of $\delta$-functions, we arrive at an expression of the form
 \begin{equation}\label{F2}
F(\{v\})= \int d{\bf k}_{i_1}\int d{\bf k}_{i_2}\int d{\bf k}_{i_3} \int d{\bf k}_{i_4}\,\frac{1}{(C+V_{i_j,i_l}{\bf k}_{i_j}{\bf k}_{i_l})^\alpha}\,,\qquad C\equiv \tau \sum_{j=1}^7 u_j.
\end{equation}
 Calculating the integral of the power of the quadratic form, we obtain
 \begin{equation}\label{F3}
F(\{v\})= \pi^{d L/2} C^{3d/2-\alpha}\frac{\Gamma(\alpha-d L/2)}{\Gamma(\alpha)} (\det V)^{-d/2},
\end{equation}
where $L$ is the number of loops in the diagram, in the case under consideration $L=4$.

 The value of the determinant $\det V$ in (\ref{F3}) does not depend on the choice of the variables of integration  $\{{\bf k}_{i_1},{\bf k}_{i_2},{\bf k}_{i_3},{\bf k}_{i_4}\}$ and can be determined directly from the diagram view. By construction, $\det V$ is the sum of products of four factors $u_i$. For any set of independent variables of integration   $\{{\bf k}_{i_1},{\bf k}_{i_2},{\bf k}_{i_3},{\bf k}_{i_4}\}$ the diagonal elements of the matrix $V$ are equal to  $u_{i_1}, u_{i_2}, u_{i_3}, u_{i_4}$, their product contributes to $\det V$ with coefficient one. The nondiagonal elements of matrix $V$ do not contain the parameters $u_{i_1}, u_{i_2}, u_{i_3}, u_{i_4}$, consequently, $\det V$ does not contain the highest powers of $u_i$.
Obviously, we can not choose as independent variables some sets of $\{{\bf k}_{i_1},{\bf k}_{i_2},{\bf k}_{i_3},{\bf k}_{i_4}\}$ which form conservation laws. As a result such a products of $u_i$ will not appear in $\det V$. 

For our particular diagram,  in addition to the sets $\{{\bf k}_{1},{\bf k}_{2},{\bf k}_{7}\}$, $\{{\bf k}_{5},{\bf k}_{6},{\bf k}_{7}\}$ and $\{{\bf k}_{1}, {\bf k}_{2}, {\bf k}_{3},{\bf k}_{4}\}$, which are defined by the $\delta$-function arguments in (\ref{J}), ``conservation laws''\, are also formed by sets
 $\{{\bf k}_{3},{\bf k}_{3},{\bf k}_{7}\}$, $\{{\bf k}_{3},{\bf k}_{4},{\bf k}_{5},{\bf k}_{6}\}$ and $\{{\bf k}_{1},{\bf k}_{2},{\bf k}_{5},{\bf k}_{6}\}$.
  Thus, out of 35 possible quadruples of products of the parameters $u_i$ 15 products do not contribute to $\det V$, and $\det V$ for the diagram (\ref{dyn2}) is given by the expression
 \begin{equation}\label{det}
 \begin{split}
\det V & = u_1u_2u_3u_5+u_1u_2u_3u_6+ u_1u_2u_4u_5+u_1u_2u_4u_6+ u_1u_3u_4u_5+  \\
&+u_1u_3u_4u_6 +u_1u_3u_5u_6+ u_1u_3u_5u_7+u_1u_3u_6u_7+u_1u_4u_5u_6+ \\
& +u_1u_4u_5u_7+ u_1u_4u_6u_7+u_2u_3u_4u_5+ u_2u_3u_4u_6 +u_2u_3u_5u_6+\\
& + u_2u_3u_5u_7+u_2u_3u_6u_7+ u_2u_4u_5u_6
+u_2u_4u_5u_7+u_2u_4u_6u_7
 \end{split}
\end{equation}
 in which $u_i$ must be expressed in terms of $v_i$ according to (\ref{uv}). For the first and third diagrams of the formula (\ref{d1=3res}), the expression (\ref{det}) is preserved, only the connections  (\ref{uv}) of the variables $u_i$ and $v_i$  will change accordingly. This can be easily found by the form of the diagram.
 

\end{document}